\documentclass{article}

\usepackage[preprint]{neurips_2024}

\usepackage[utf8]{inputenc}
\usepackage[T1]{fontenc}
\usepackage{hyperref}
\usepackage{url}
\usepackage{booktabs}
\usepackage{amsfonts}
\usepackage{nicefrac}
\usepackage{microtype}
\usepackage{graphicx}
\usepackage{adjustbox}
\usepackage{array}
\usepackage{arydshln}
\usepackage{subcaption}
\usepackage{amsmath}
\usepackage{amssymb}
\usepackage{multirow}
\usepackage[table]{xcolor}
\usepackage{cleveref}
\usepackage{xspace}
\usepackage{placeins}
\usepackage{float}
\usepackage{tikz}
\usetikzlibrary{arrows.meta,positioning,calc}

\newcommand{\ours}{F3-Tokenizer\xspace}
\definecolor{aeBlue}{HTML}{2F6FB0}
\definecolor{semGreen}{HTML}{2F8F5B}
\definecolor{genOrange}{HTML}{C97822}
\definecolor{lightGray}{HTML}{F6F7F9}
\definecolor{panelGray}{HTML}{F1F3F5}
\definecolor{tokenBlue}{HTML}{62A9F5}
\definecolor{tokenTeal}{HTML}{8ACBC6}
\definecolor{headPurple}{HTML}{D8D5FA}
\definecolor{lmGold}{HTML}{FFDFA6}

\title{F3-Tokenizer: Taming Audio Autoencoder Latents for\\
Understanding and Generation}

\author{%
  Dinghao Zhou$^{2,*}$ \quad
  Xingchen Song$^{2,*}$ \quad
  Di Wu$^{2,*}$ \quad
  Pengyu Cheng$^{*}$\\
  Shengfan Shen$^{1,*}$ \quad
  Sixiang Lv$^{1,*}$\\
  \vspace{0.15cm}\\
  $^1$Nanjing University, China \quad
  $^2$WeNet Open Source Community\\
  $^*$Equal contribution
}

\begin{document}

\maketitle

\begin{abstract}
Continuous audio autoencoders reconstruct waveforms well but often produce
latents with weak structure for understanding, while self-supervised audio encoders
capture semantics but are not directly decodable. This mismatch complicates a
single audio tokenizer that must support both understanding and generation. We
adapt continuous autoencoder latents to this setting with two components: a
noise-regularized autoencoder bottleneck and a latent-side representation
encoder. The bottleneck uses channel normalization and stochastic perturbation
instead of KL-based variational training, yielding scale-controlled continuous
latents for reconstruction and autoregressive generation. The representation
encoder is trained on frozen autoencoder latents with RQ-MTP and frozen-LLM
supervision. The resulting tokenizer provides high-dimensional representations
for understanding while preserving normalized continuous latents as generation
targets.
\end{abstract}

\section{Introduction}

Audio representation learning sits at the center of recent progress in speech
language models, text-to-speech synthesis, and general audio generation. A
common strategy is to compress waveform into a sequence of acoustic latents and
then train a language-conditioned generator over that sequence~\cite{audiolm,
valle,latentlm,vibevoice}. The quality of the latent space therefore determines
what the resulting generator can do: preserve intelligible speech, retain speaker
and acoustic conditions, support long-form synthesis, and remain controllable by
text or other high-level conditions.

Existing audio representations tend to specialize in only part of this problem.
Neural codecs and audio autoencoders are optimized for reconstruction and
compression, producing latents that retain acoustic detail but may be poorly
organized for audio understanding~\cite{soundstream,encodec}.
Self-supervised audio encoders, in contrast, learn representations useful for
recognition and classification~\cite{wav2vec2,hubert,nestrq}, but they are not
designed to be decoded into waveform. Recent continuous audio generators use
patch-wise diffusion or flow heads over latent targets~\cite{flowmatching,voicebox,
latentlm,ditar,vibevoice,semavoice}, yet the target latent is often inherited from a
reconstruction model and treated as fixed.

Existing systems often improve semantic structure by adding another
representation path. X-Codec-2.0 fuses a pretrained semantic encoder with an
acoustic encoder before quantization~\cite{xcodec2}; VibeVoice uses separate
acoustic and semantic tokenizers for long-form continuous speech
generation~\cite{vibevoice}; MiMo-Audio-Tokenizer trains a large discrete
tokenizer with both semantic and reconstruction objectives~\cite{mimoaudio};
and MingTok-Audio explores a continuous speech tokenizer with
Whisper-distilled semantics~\cite{minguniaudio}. These designs confirm the
value of semantic information, but they can introduce separate encoders,
external teachers, discrete codebooks, or mismatched representation formats
between understanding representations, decoder inputs, and continuous generation
targets. We keep the continuous autoencoder latent as the acoustic anchor: the
low-dimensional latent remains the reconstructive and generative target, while
a representation encoder built on top of it provides high-dimensional representations
for understanding.

Our focus is the tokenizer itself. We build on a SpectroStream-style
STFT-domain audio autoencoder~\cite{spectrostream}, but replace its bottleneck
with a normalized, noise-regularized continuous latent inspired by
continuous-token autoregressive generation work~\cite{nextstep1}. This produces
normalized continuous latents without KL regularization, keeping the autoencoder
latent decodable and stable as an acoustic target for reconstruction and
generation. On top of frozen autoencoder latents, we train a latent-side
representation encoder with RQ-MTP self-supervision and supervision from a
frozen LLM. During tokenizer training, projected audio embeddings enter the
frozen LLM, and the resulting audio-aligned states condition a next-patch flow
head that predicts continuous autoencoder latent patches.

The scope is tokenizer design rather than a full unified audio-language model. We study
whether normalized autoencoder latents, representation learning, and
generation-side flow supervision provide three properties: \textbf{acoustic
fidelity}, \textbf{understanding utility}, and \textbf{predictability for
audio generation}.
We refer to the resulting system as \ours, emphasizing fidelity,
high-dimensional representations, and flow-based generation.

Our contributions are:
\begin{itemize}
    \item We formulate audio autoencoder latents as acoustic anchors for a
    tokenizer pipeline that exposes complementary outputs for reconstruction,
    understanding, and generation.
    \item We use a normalized continuous bottleneck with per-token channel
    normalization and uniform-strength stochastic perturbation, avoiding a
    VAE-style KL objective while improving robustness of the latent space.
    \item We attach a representation encoder trained with
    random-quantized multi-token prediction and frozen-LLM supervision,
    producing high-dimensional representations on top of frozen autoencoder
    latents.
    \item We co-train a generation-side patch-level flow head that maps
    frozen-LLM states conditioned on downsampled representations to continuous
    autoencoder latent patches.
\end{itemize}

\section{Related Work}

\paragraph{Neural audio codecs and continuous autoencoders.}
Neural audio codecs compress waveform into compact latent sequences and
reconstruct audio through a decoder. SoundStream~\cite{soundstream} introduced
an end-to-end codec with residual vector quantization, and EnCodec~\cite{encodec}
demonstrated high-fidelity real-time audio compression. These reconstruction
modules have become common tokenizers for speech and audio language models.
More recent work has also moved toward continuous autoencoder targets for
generative modeling. Continuous latent tokenizers often use VAE-style KL
regularization to shape the latent distribution. LatentLM~\cite{latentlm}
further shows that autoregressive modeling over continuous latents requires
careful variance control, introducing a $\sigma$-VAE variant to mitigate
variance collapse. These works suggest that continuous latents should be
distribution-controlled for generation, while leaving open whether such control
must come from a variational posterior.

\paragraph{Semantic-acoustic tokenizer fusion.}
A common strategy for improving tokenizer semantics is to add a semantic path
beside the acoustic reconstruction path. AudioLM~\cite{audiolm} explicitly
separates semantic and acoustic tokens for long-range audio continuation.
X-Codec~\cite{xcodec} and X-Codec-2.0~\cite{xcodec2} inject pretrained
semantic features into codec training before quantization and add semantic
reconstruction losses, improving codec tokens for audio language modeling.
MiMo-Audio-Tokenizer~\cite{mimoaudio} jointly optimizes semantic understanding
and acoustic reconstruction in a large discrete RVQ tokenizer.
MingTok-Audio~\cite{minguniaudio} moves closer to continuous-token designs, but obtains
speech semantics through a Whisper-initialized module and distillation from a
frozen Whisper encoder, which can bias the tokenizer toward ASR-centric
attributes. VibeVoice~\cite{vibevoice} uses separate acoustic and semantic
tokenizers, while SemaVoice~\cite{semavoice} refines continuous speech
representations with a speech-foundation-model guided alignment mechanism.
These designs highlight the importance of semantic-acoustic fusion, but often
introduce extra teachers, separate encoders, discrete codebooks, or multiple
token streams. We keep the continuous autoencoder latent as the decodable
acoustic target and learn high-dimensional representations on top of it.

\paragraph{Self-supervised audio representation learning.}
Self-supervised models such as wav2vec 2.0~\cite{wav2vec2} and
HuBERT~\cite{hubert} learn representations from unlabeled speech that transfer
well to recognition tasks. NEST-RQ~\cite{nestrq} further studies causal
next-token prediction with random-projection quantization. Such self-supervised training is
effective for understanding, but the resulting representations are not usually
trained to decode high-fidelity waveform. Our representation encoder borrows this
RQ-style self-supervision idea, while keeping it attached to a reconstructive
autoencoder latent and decoder-facing projection.

\paragraph{Autoregressive continuous audio generation.}
Flow matching~\cite{flowmatching} provides a simulation-free objective for
learning continuous generative models, and Voicebox~\cite{voicebox} applies
flow matching to text-guided speech generation. Recent audio language models
increasingly generate continuous representations with diffusion or flow heads
rather than only discrete codec tokens. LatentLM~\cite{latentlm} studies
next-token diffusion over continuous VAE latents and introduces a $\sigma$-VAE
variant to mitigate variance collapse in autoregressive latent modeling, while
DiTAR~\cite{ditar}, VibeVoice~\cite{vibevoice}, and
SemaVoice~\cite{semavoice} use patch-wise diffusion or flow heads over
continuous speech representations. DiTAR, for example, combines a language
model over patch embeddings with a diffusion transformer that generates the
next continuous speech patch. We use this next-patch view as tokenizer
supervision: during tokenizer training, a flow head conditioned on frozen-LLM
states predicts continuous autoencoder latent patches. This makes the target
space generation-aware while keeping the paper focused on the tokenizer rather
than a complete unified audio-language model.

\section{Method}

\subsection{Normalized Autoencoder}

Our autoencoder follows a SpectroStream-style STFT-domain backbone, but
replaces the original bottleneck with a normalized continuous latent. Given an
audio waveform $x$, we let $E_\phi$ include the STFT frontend and audio encoder:
it maps $x$ to hidden states $h=E_\phi(x)$, followed by a raw bottleneck
$z_0=B_\phi(h)$ with latent dimension $D=64$ in our experiments. We then
normalize each time step across channels:
\begin{equation}
z_n = \frac{z_0 - \mu_c(z_0)}
{\sigma_c(z_0) + \epsilon_{\mathrm{std}}},
\end{equation}
where $\mu_c$ and $\sigma_c$ are computed over the channel dimension. During
training, we add controlled stochastic perturbation:
\begin{equation}
\tilde{z} = z_n + \alpha \epsilon,\qquad
\alpha \sim \mathcal{U}(0,\gamma),\quad
\epsilon \sim \mathcal{N}(0,I).
\end{equation}
The decoder reconstructs waveform from the perturbed latent during training,
\begin{equation}
\hat{x}=D_\psi(\tilde{z}),
\end{equation}
and is optimized with spectral reconstruction losses plus GAN-style
adversarial losses, following the audio reconstruction training recipes of
DAC~\cite{dac} and SpectroStream~\cite{spectrostream}. Deterministic
encoding uses $z_n$ directly. In the rest of the paper, we
write $z$ for this normalized autoencoder latent. The latent $z$ remains the
decodable acoustic target for reconstruction and generation; a representation
encoder later turns it into high-dimensional, LLM-compatible representations
for understanding.

\begin{figure}[t]
\centering
\makebox[\textwidth][c]{%
\resizebox{\textwidth}{!}{%
\begin{tikzpicture}[
    font=\sffamily\small,
    >=Latex,
    panel/.style={fill=panelGray, rounded corners=8pt},
    title/.style={font=\sffamily\bfseries\large, align=center},
    module/.style={draw=none, rounded corners=6pt, minimum height=0.66cm,
        align=center, font=\sffamily\bfseries\small},
    smallmod/.style={draw=none, rounded corners=5pt, minimum height=0.50cm,
        align=center, font=\sffamily\footnotesize},
    tag/.style={draw=black!20, rounded corners=3pt, line width=0.4pt,
        align=center, inner xsep=4pt, inner ysep=2pt,
        font=\sffamily\scriptsize, fill=white},
    ztok/.style={draw=tokenTeal!85!black, rounded corners=2pt, line width=0.45pt,
        minimum width=0.34cm, minimum height=0.34cm, fill=tokenTeal!42},
    utok/.style={draw=semGreen!70!black, rounded corners=2pt, line width=0.45pt,
        minimum width=0.34cm, minimum height=0.34cm, fill=semGreen!35},
    ttok/.style={draw=tokenBlue!70, rounded corners=2pt, line width=0.45pt,
        minimum width=0.34cm, minimum height=0.34cm, fill=tokenBlue!35},
    flowtok/.style={draw=genOrange!65, dashed, rounded corners=2pt,
        line width=0.45pt, minimum width=0.34cm, minimum height=0.34cm,
        fill=semGreen!18},
    tallz/.style={ztok, rounded corners=3pt, minimum width=0.26cm,
        minimum height=0.64cm},
    tallu/.style={utok, rounded corners=3pt, minimum width=0.26cm,
        minimum height=0.64cm},
    tallt/.style={ttok, rounded corners=3pt, minimum width=0.26cm,
        minimum height=0.64cm},
    arr/.style={->, line width=0.85pt, draw=black!55},
    aux/.style={->, dashed, line width=0.70pt},
    count/.style={font=\sffamily\scriptsize, text=black!55, inner sep=0pt},
]
\path[panel] (0,0) rectangle (4.40,6.25);
\path[panel] (4.75,0) rectangle (14.10,6.25);

\node[title] at (2.20,6.66) {F3-Tokenizer};
\node[title] at (9.43,6.66) {\normalsize Representation and Flow Training};

\coordinate (inputX) at (2.20,4.98);
\draw[line width=0.78pt, draw=aeBlue]
    (1.58,5.24) .. controls (1.72,5.40) and (1.86,5.08) .. (2.00,5.24)
    .. controls (2.14,5.40) and (2.28,5.08) .. (2.42,5.24)
    .. controls (2.56,5.40) and (2.70,5.08) .. (2.84,5.24);
\path[draw=aeBlue!55, fill=aeBlue!18, rounded corners=5pt, line width=0.5pt]
    (0.78,4.72) -- (3.62,4.72) -- (3.28,3.88) -- (1.12,3.88) -- cycle;
\node[font=\sffamily\bfseries\small, align=center] (aeEncText) at (2.20,4.30)
    {AE Encoder};
\foreach \x in {1.48,1.84,2.20,2.56,2.92}
    \node[tallz] at (\x,3.24) {};
\node[tag, draw=aeBlue!40] at (3.45,3.58) {norm + noise};
\node[module, fill=aeBlue!18, minimum width=1.40cm, minimum height=0.70cm,
    font=\sffamily\bfseries\scriptsize]
    (aeDec) at (1.38,1.84) {AE Decoder};
\node[module, fill=headPurple, minimum width=1.86cm, minimum height=0.70cm,
    font=\sffamily\bfseries\tiny]
    (semTok) at (3.10,1.84) {Representation\\Encoder};
\draw[line width=0.78pt, draw=aeBlue]
    (0.86,0.76) .. controls (0.98,0.88) and (1.10,0.64) .. (1.22,0.76)
    .. controls (1.34,0.88) and (1.46,0.64) .. (1.58,0.76)
    .. controls (1.70,0.88) and (1.82,0.64) .. (1.94,0.76);
\node[font=\sffamily\scriptsize] at (1.38,0.34) {$\hat{x}$};
\foreach \x in {2.36,2.73,3.10,3.47,3.84}
    \node[tallu] at (\x,0.78) {};
\draw[arr] (inputX) -- (2.20,4.72);
\draw[arr] (2.20,3.88) -- (2.20,3.62);
\draw[arr] (1.96,2.86) -- (aeDec.north);
\draw[arr] (2.44,2.86) -- (semTok.north);
\draw[arr] (aeDec.south) -- (1.38,1.06);
\draw[arr] (semTok.south) -- (3.05,1.12);

\foreach \x in {7.60,8.10,8.60,9.10,9.60,10.10,10.60,11.10}
    \node[tallz] at (\x,0.42) {};
\foreach \bl/\br in {7.42/8.28,8.42/9.28,9.42/10.28,10.42/11.28} {
    \draw[dashed, draw=black!24, rounded corners=3pt]
        (\bl,0.04) rectangle (\br,0.80);
}

\node[module, fill=headPurple, minimum width=4.70cm, minimum height=0.54cm]
    (semTrain) at (9.35,1.18) {Representation Encoder};
\foreach \x in {7.60,8.10,8.60,9.10,9.60,10.10,10.60,11.10}
    \node[tallu] at (\x,1.88) {};
\foreach \bl/\br in {7.42/8.28,8.42/9.28,9.42/10.28,10.42/11.28} {
    \draw[dashed, draw=black!24, rounded corners=3pt]
        (\bl,1.50) rectangle (\br,2.26);
}
\foreach \c in {7.85,8.85,9.85,10.85} {
    \draw[arr, draw=semGreen!55!black, line width=0.55pt] (\c,2.22) -- (\c,2.86);
}

\node[smallmod, fill=white, minimum width=3.90cm, minimum height=0.42cm,
    font=\sffamily\scriptsize]
    (downProj) at (9.35,2.52) {downsample + project};
\foreach \x in {7.85,8.85,9.85,10.85}
    \node[tallu] at (\x,3.24) {};

\foreach \x in {6.00,6.45,6.90,7.35}
    \node[tallt] at (\x,3.24) {};

\node[module, fill=lmGold, minimum width=7.10cm, minimum height=0.76cm]
    (lmTrain) at (9.35,4.12) {LLM};

\foreach \c in {7.85,8.85,9.85,10.85} {
    \node[module, fill=headPurple, minimum width=0.82cm, minimum height=0.62cm,
        font=\sffamily\bfseries\tiny]
        at (\c,4.86) {Patch\\DiT};
}
\foreach \x in {7.60,8.10,8.60,9.10,9.60,10.10,10.60,11.10}
    \node[tallz] at (\x,5.66) {};
\foreach \bl/\br in {7.42/8.28,8.42/9.28,9.42/10.28,10.42/11.28} {
    \draw[dashed, draw=black!24, rounded corners=3pt]
        (\bl,5.28) rectangle (\br,6.04);
}
\draw[line width=0.58pt, draw=aeBlue]
    (2.10,-0.28) .. controls (2.18,-0.20) and (2.26,-0.36) .. (2.34,-0.28)
    .. controls (2.42,-0.20) and (2.50,-0.36) .. (2.58,-0.28);
\node[font=\sffamily\tiny, text=black!62, anchor=west] at (2.74,-0.28) {waveform};
\node[tallt, minimum width=0.16cm, minimum height=0.30cm] at (4.42,-0.28) {};
\node[font=\sffamily\tiny, text=black!62, anchor=west] at (4.60,-0.28) {text token};
\node[tallu, minimum width=0.16cm, minimum height=0.30cm] at (6.54,-0.28) {};
\node[font=\sffamily\tiny, text=black!62, anchor=west] at (6.72,-0.28) {representation};
\node[tallz, minimum width=0.16cm, minimum height=0.30cm] at (8.92,-0.28) {};
\node[font=\sffamily\tiny, text=black!62, anchor=west] at (9.10,-0.28) {acoustic latent};

\end{tikzpicture}%
}}
\caption{Overview of the tokenizer components and training pipeline. The left
panel shows the tokenizer outputs: an autoencoder exposes normalized
continuous latents $z$ for reconstruction, while a representation encoder
produces high-dimensional representations $u$. The right panel shows how the representation
encoder is trained:
$u$ is patch-projected into audio tokens for a frozen LLM, and a patch DiT head
is trained to predict continuous $z$ patches.}
\label{fig:method}
\end{figure}
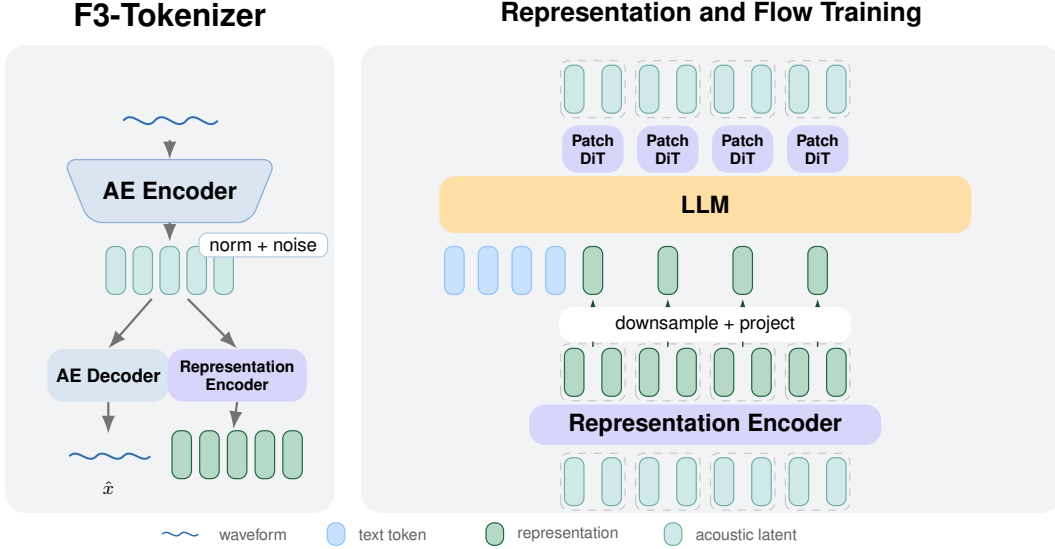

\subsection{Latent-Side Representation Encoder}

The autoencoder latent $z$ is passed to a causal representation encoder
$R_\eta$ with a projection back to the decoder dimension. For long audio, we
apply the encoder with a sliding-window schedule so that training remains
causal without full-sequence computation:
\begin{equation}
s, u = R_\eta(z).
\end{equation}
The representation sequence $u$ is used for self-supervised learning and
language-grounded supervision. A decoder-side projection $s$ ties the
representation path back to acoustic reconstruction while leaving $z$ as the
continuous target for generation.

\paragraph{RQ-MTP and LLM supervision.}

We train prediction heads over $u$ with random-quantized multi-token prediction
(RQ-MTP), following the next-token prediction formulation of
NEST-RQ~\cite{nestrq}. The main head predicts the next quantized target and
auxiliary heads can predict additional future targets:
\begin{equation}
\mathcal{L}_{\mathrm{rq\text{-}mtp}}
= \sum_{k=1}^{K}
\mathrm{CE}\left(g_k(u_t), c_{t+k}\right),
\end{equation}
where $c_{t+k}$ denotes the random-quantized target at a future frame. RQ-MTP
encourages temporal structure useful for understanding in the latent-side
representation while remaining compatible with continuous autoencoder latents.

When paired text is available, the representations $u$ are temporally
downsampled and projected as audio embeddings before being fed to a frozen LLM:
\begin{equation}
e = P_\rho(\mathrm{Down}(u)).
\end{equation}
The LLM weights remain fixed; the text cross-entropy loss updates
the representation encoder and its projection into the LLM input space:
\begin{equation}
\mathcal{L}_{\mathrm{lm}} = \mathrm{CE}(p_\omega(y \mid e), y).
\end{equation}
We denote by $q_t$ the audio-aligned hidden state of this frozen LLM
at audio patch $t$.

\paragraph{Reconstruction constraint.}

The projected sequence $s$ is tied back to waveform through the autoencoder
decoder, so the representation encoder cannot discard acoustic information. We
preserve acoustic detail with the decoder reconstruction objective:
\begin{equation}
\mathcal{L}_{\mathrm{recon}}
= \mathcal{L}_{\mathrm{ae}}(x,\hat{x}).
\end{equation}
This constraint keeps the representation path tied to waveform-relevant
acoustic information.

\subsection{Patch-Wise Flow Head}

For generation, the continuous autoencoder latent is modeled patch by patch.
Let $P$ be the number of frame-level latents represented by one autoregressive
audio-token step.
For patch index $t$, the continuous target is
\begin{equation}
a_t =
\left[z_{tP}, z_{tP+1}, \ldots, z_{(t+1)P-1}\right],
\end{equation}
where $z$ is the normalized autoencoder latent. During Stage~1, a frozen LLM
consumes text tokens and projected audio tokens derived from $u$; we denote by
$q_t$ the hidden state aligned to audio patch $t$. A patch-wise flow head
conditions on $q_t$, optionally together with previous $z$ patches as acoustic
history, and predicts the whole continuous patch $a_t$ in one autoregressive
step. This follows the next-patch formulation used by continuous audio
generators such as DiTAR~\cite{ditar}, while keeping $z$ as the
decoder-facing acoustic target. The head is trained with a standard
flow-matching objective~\cite{flowmatching} over each target patch.

\subsection{Training Pipeline}

Figure~\ref{fig:method} summarizes the tokenizer outputs and training
pipeline. We use a staged recipe: first learn a decodable autoencoder latent,
then train the representation encoder and generation-side flow head, and
finally evaluate the learned representations on task-specific systems.

\paragraph{Stage 0: normalized autoencoder training.}
The first stage trains the autoencoder with the normalized and noise-perturbed
bottleneck described in Section~3.1. This stage optimizes waveform
reconstruction and exposes the continuous latent $z$ as both the decoder input
and the default generation target.

\paragraph{Stage 1: representation and flow training.}
We freeze the autoencoder and the LLM, and train only the representation
encoder, the projection into the LLM input space, and the generation-side patch
flow head. Audio-only data provides random-quantized multi-token prediction on
$u$. Audio-text pairs, including ASR transcripts and audio captions, provide
frozen-LLM supervision after $u$ is downsampled and projected into audio tokens.
All audio examples also provide a next-patch flow loss: an audio-aligned hidden
state from the frozen LLM conditions a patch head that predicts the next
continuous $z$ patch. For paired examples, the frozen LLM can additionally
condition on text tokens; for audio-only examples, it conditions on audio-token
history.

\paragraph{Stage 2: task-specific training.}
Recent work on unified audio-language models often aims to support
understanding and generation in one backbone. We do not train such a unified
model; the paper isolates the tokenizer. We therefore evaluate the learned
tokenizer through separate tasks such as ASR, TTS, and text-to-audio (TTA). In
these experiments, the task models are trainable while the tokenizer supplies
the acoustic representations. Understanding tasks use the high-dimensional
representation $u$ for task-specific prediction. Generation tasks train an
autoregressive audio generator over audio tokens derived from $u$; its patch
flow head predicts continuous $z$ patches decoded by the autoencoder decoder.

\section{Experiments}

We evaluate the tokenizer from three perspectives: reconstruction,
understanding, and generation.

\subsection{Reconstruction Quality}

We measure whether the normalized autoencoder latent preserves acoustic fidelity
relative to continuous autoencoder baselines. Table~\ref{tab:recon} compares a
VibeVoice-style $\sigma$-VAE tokenizer with our autoencoder variants and
isolates the effect of normalization and noise injection in the bottleneck.

\begin{table}[htbp]
\centering
\caption{Reconstruction and preservation metrics across speech, music, and
sound. We compare against VibeVoice and ablate normalization and noise injection
in the autoencoder bottleneck. Speech metrics are evaluated on AISHELL-3
(24~kHz Chinese) and LibriTTS test-other (24~kHz English). Music metrics are evaluated on
MUSDB18-HQ (24~kHz). General audio metrics are
evaluated on AudioCaps. Slash-separated entries, where present, follow the
speech dataset order above.}
\label{tab:recon}
\resizebox{\textwidth}{!}{%
\begin{tabular}{lccccccccccc}
\toprule
\multirow{2}{*}{Model} & \multirow{2}{*}{Token Rate}
& \multicolumn{5}{c}{Speech}
& \multicolumn{3}{c}{Music}
& \multicolumn{2}{c}{Sound} \\
\cmidrule(lr){3-7}\cmidrule(lr){8-10}\cmidrule(lr){11-12}
& & MCD$\downarrow$ & M-STFT$\downarrow$ & PESQ$\uparrow$ & STOI$\uparrow$ & ViSQOL$\uparrow$
& KL$_{\mathrm{PaSST}}\downarrow$ & FD$_{\mathrm{OpenL3}}\downarrow$ & ViSQOL$\uparrow$
& KL$_{\mathrm{PaSST}}\downarrow$ & FD$_{\mathrm{OpenL3}}\downarrow$ \\
\midrule
VibeVoice ($\sigma$-VAE) & 7.5 Hz & 5.19/3.90 & 2.88/1.26 & 2.93/3.01 & 0.935/0.941 & 4.39/4.31 & 0.0580 & 150.99 & 3.68 & 0.6540 & 51.10 \\
Autoencoder & 25 Hz & 2.58/3.41 & 1.74/1.08 & 2.96/\textbf{3.22} & 0.932/0.918 & 4.32/4.38 & 0.0308 & 34.20 & 4.32 & 0.4650 & 31.40 \\
Autoencoder w/ norm + noise & 25 Hz & \textbf{2.33/3.27} & \textbf{1.61/0.99} & \textbf{3.07}/3.21 & \textbf{0.940/0.946} & \textbf{4.41/4.68} & \textbf{0.0269} & \textbf{29.02} & \textbf{4.44} & \textbf{0.2164} & \textbf{15.24} \\
\bottomrule
\end{tabular}
}
\end{table}
\FloatBarrier

\subsection{Audio Understanding}

We test whether the learned representation is useful for audio understanding
with a frozen-feature probing protocol. The autoencoder and representation
encoder are fixed, and only task-specific probes are trained on top of the
extracted representation. This keeps the evaluation focused on the representation
rather than the capacity of a large task model. As a low-information control,
we also train the same probes without the learned representation, so they can
rely only on dataset priors.
Table~\ref{tab:understanding} reports the probing results and ablations.
All scores are reported as percentages and higher is better.

The results should be read within each domain rather than averaged across
datasets, since task sizes and label spaces differ. Removing RQ-MTP weakens the
audio-only self-supervised signal and causes broad drops on speech, sound, and
music tasks. Removing frozen-LLM supervision mainly hurts speech-language tasks
such as FSC, LibriSpeech-100h, and Speech Commands, indicating that language
alignment complements predictive self-supervision. Combining both objectives
yields the strongest representation across domains.

\paragraph{Probe-task summary.}
The probing suite spans speech, environmental sound, and music understanding
tasks; the compact table below lists the dataset groups used in
Table~\ref{tab:understanding}.
\begin{center}
\vspace{-0.4em}
\scriptsize
\setlength{\tabcolsep}{3.0pt}
\resizebox{\textwidth}{!}{%
\begin{tabular}{llll}
\toprule
Domain & Dataset & Task & Prediction target \\
\midrule
Speech & ASV2015 & speaker verification & trial label \\
Speech & CREMA-D, RAVDESS & emotion recognition & emotion class \\
Speech & FSC & spoken command understanding & intent / action class \\
Speech & LibriCount & speaker counting & number of speakers \\
Speech & LibriSpeech-100h & speech recognition probing & transcript-related label \\
Speech & LibriSpeech-MF & speaker attribute probing & gender label \\
Speech & Speech Cmds V1 & keyword spotting & command class \\
Speech & Vocal Imitation, VocalSound & vocal sound recognition & vocal category \\
Speech & VoxCeleb1 & speaker identification & speaker identity \\
Sound & DESED & sound event detection & event label \\
Sound & ESC-50, UrbanSound8K & environmental sound classification & scene / sound class \\
Sound & FSD50K, FSD18-Kaggle & general audio tagging & sound-event tag \\
Music & FMA Small, GTZAN & music genre classification & genre label \\
Music & NSynth & instrument recognition & instrument family \\
\bottomrule
\end{tabular}
}
\vspace{-0.6em}
\end{center}

\begin{table}[!htbp]
\centering
\caption{Frozen-representation probing on speech, sound, and music
understanding datasets. ``Ming-U'' denotes MingTok-Audio (Unified). ``No
repr.'' removes the learned representation, ``w/o RQ'' removes RQ-MTP, ``w/o
LLM'' removes frozen-LLM supervision, and \ours{} uses all objectives.}
\label{tab:understanding}
\scriptsize
\setlength{\tabcolsep}{2.5pt}
\setlength{\dashlinedash}{1.0pt}
\setlength{\dashlinegap}{1.4pt}
\resizebox{\textwidth}{!}{%
\begin{tabular}{llcc:cccc}
\toprule
& & \multicolumn{2}{c}{Baselines} & \multicolumn{4}{c}{Ours} \\
\cmidrule(lr){3-4}\cmidrule(lr){5-8}
Domain & Dataset & Whisper & Ming-U & No repr. & w/o RQ & w/o LLM & \ours{} \\
\midrule
\multirow{11}{*}{Speech}
& ASV2015 & 96.60 & 98.70 & 49.80 & 94.70 & 96.10 & \textbf{99.65} \\
& CREMA-D & 57.20 & 66.85 & 12.50 & 61.20 & 70.40 & \textbf{78.90} \\
& FSC & 77.60 & \textbf{98.58} & 1.10 & 88.60 & 84.20 & 94.86 \\
& LibriCount & 54.90 & 62.92 & 9.80 & 62.40 & 68.00 & \textbf{76.10} \\
& LibriSpeech-100h & 81.50 & 93.45 & 0.00 & 86.40 & 80.80 & \textbf{96.00} \\
& LibriSpeech-MF & 97.30 & 97.10 & 50.20 & 94.50 & 95.40 & \textbf{98.40} \\
& RAVDESS & 45.90 & 57.08 & 11.40 & 55.20 & 61.10 & \textbf{70.10} \\
& Speech Cmds V1 & 93.30 & 96.14 & 8.70 & 88.80 & 84.60 & \textbf{96.80} \\
& Vocal Imitation & 18.00 & 20.64 & 1.60 & 16.80 & 19.20 & \textbf{21.90} \\
& VocalSound & 86.00 & 88.59 & 20.50 & 80.30 & 86.20 & \textbf{91.60} \\
& VoxCeleb1 & 38.80 & 35.28 & 3.20 & 56.80 & 65.40 & \textbf{72.00} \\
\midrule
\multirow{5}{*}{Sound}
& DESED & 12.70 & 39.80 & 9.40 & 34.20 & 43.20 & \textbf{50.10} \\
& ESC-50 & 52.80 & 62.25 & 12.70 & 57.80 & 69.50 & \textbf{75.20} \\
& FSD50K & 26.20 & 24.83 & 2.60 & 21.20 & 28.70 & \textbf{32.10} \\
& FSD18-Kaggle & 24.10 & 40.81 & 6.10 & 63.50 & 75.40 & \textbf{81.38} \\
& UrbanSound8K & 68.70 & 72.87 & 28.50 & 62.80 & 70.60 & \textbf{74.22} \\
\midrule
\multirow{3}{*}{Music}
& FMA Small & 58.10 & 54.45 & 21.60 & 54.40 & 60.50 & \textbf{64.84} \\
& GTZAN & 62.20 & 71.17 & 14.20 & 72.60 & 80.40 & \textbf{85.00} \\
& NSynth & 53.20 & 57.00 & 16.40 & 50.10 & 54.60 & \textbf{58.12} \\
\bottomrule
\end{tabular}
}
\end{table}
\FloatBarrier

\subsection{Audio Generation}

We evaluate generation by training autoregressive audio generators on top of
the proposed tokenizer outputs. In both settings, the representation encoder
produces high-dimensional frame-level representations $u$, a patch projection
converts them into audio tokens, and a patch flow head maps generator hidden
states back to continuous $z$ patches for waveform decoding.

\paragraph{Text-to-speech.}
For TTS, we train speech generators and evaluate them on the Seed-zh and
Seed-en test sets. The token-rate column in
Table~\ref{tab:generation} reports the conversion from the 25~Hz autoencoder
latent stream to the audio-token stream consumed by the generator.

\paragraph{Text-to-audio.}
For AudioCaps TTA, we use a two-step downstream schedule. Before training on
AudioCaps, we warm up the patch projection and autoregressive backbone on several
thousand hours of audio-caption pairs with a captioning objective, so that the
model first learns audio-text alignment. We then train the text-to-audio
generator on AudioCaps using the same tokenizer outputs and patch-flow target.

\begin{table}[!htbp]
\centering
\caption{Generation results for text-to-speech and text-to-audio. CER/WER is
reported in percent, and SIM denotes speaker similarity.}
\label{tab:generation}
\scriptsize
\setlength{\tabcolsep}{3.5pt}
\textbf{Text-to-speech on Seed-zh and Seed-en}\\[0.3em]
\resizebox{\textwidth}{!}{%
\begin{tabular}{lccccc}
\toprule
Model & Token rate & Seed-zh CER$\downarrow$ & Seed-zh SIM$\uparrow$ & Seed-en WER$\downarrow$ & Seed-en SIM$\uparrow$ \\
\midrule
CosyVoice 3-1.5B & 25 Hz & 1.12 & \textbf{0.781} & 2.21 & \textbf{0.720} \\
Qwen3-TTS-25Hz-0.6B-Base & 25 Hz & 1.18 & -- & 1.64 & -- \\
Qwen3-TTS-25Hz-1.7B-Base & 25 Hz & 1.10 & -- & 1.49 & -- \\
Qwen3-TTS-12Hz-0.6B-Base & 12 Hz & 0.92 & -- & 1.32 & -- \\
Qwen3-TTS-12Hz-1.7B-Base & 12 Hz & \textbf{0.77} & -- & \textbf{1.24} & -- \\
Ming-Flash-Omni-preview & 12.5$\rightarrow$3.1 Hz & 0.99 & 0.740 & 1.59 & 0.680 \\
Ming-omni-tts-0.5B & 12.5$\rightarrow$3.1 Hz & 0.87 & 0.72 & 2.19 & 0.61 \\
Ming-omni-tts-16.8B-A3B & 12.5$\rightarrow$3.1 Hz & 0.83 & 0.75 & 2.02 & 0.62 \\
Ming-UniAudio-16B-A3B & 50$\rightarrow$10 Hz & 0.95 & 0.70 & 1.85 & 0.58 \\
VibeVoice-1.5B & 7.5 Hz & 1.16 & 0.744 & 3.04 & 0.689 \\
\midrule
F3-Tokenizer-LLM (4B) & 25$\rightarrow$12.5 Hz & 0.90 & 0.76 & 1.88 & 0.68 \\
\bottomrule
\end{tabular}
}

\vspace{0.6em}
\textbf{Text-to-audio on AudioCaps}\\[0.3em]
\begin{tabular}{lcccc}
\toprule
Model & Token rate & FD$_{\mathrm{OpenL3}}\downarrow$ & KL$_{\mathrm{PaSST}}\downarrow$ & CLAP score$\uparrow$ \\
\midrule
Ming-omni-tts-0.5B & 12.5$\rightarrow$3.1 Hz & 74.292 & 2.257 & 0.347 \\
Ming-omni-tts-16.8B-A3B & 12.5$\rightarrow$3.1 Hz & 65.918 & 1.640 & 0.424 \\
F3-Tokenizer-LLM (4B) & 25$\rightarrow$12.5 Hz & \textbf{62.700} & \textbf{1.520} & \textbf{0.438} \\
\bottomrule
\end{tabular}
\end{table}

\begin{figure}[!htbp]
\centering
\resizebox{0.62\textwidth}{!}{%
\begin{tikzpicture}[
    font=\sffamily\small,
    panel/.style={fill=panelGray, rounded corners=8pt},
    note/.style={fill=white, rounded corners=4pt, align=center, inner sep=3pt},
    tag/.style={fill=white, draw=black!18, rounded corners=4pt, align=center,
        inner xsep=5pt, inner ysep=3pt},
    line/.style={line width=1.25pt},
    marker/.style={circle, inner sep=1.4pt},
]
\node[panel, minimum width=10.2cm, minimum height=5.3cm, anchor=south west] at (0,0) {};

\draw[->, line width=0.8pt, draw=black!55] (1.05,1.05) -- (1.05,4.85);
\draw[->, line width=0.8pt, draw=black!55] (1.05,1.05) -- (8.05,1.05);
\node[rotate=90, font=\scriptsize] at (0.35,3.0) {WER (\%) $\downarrow$};
\node[font=\scriptsize] at (4.55,0.45) {TTS training updates};
\foreach \y/\lab in {1.05/0,1.75/20,2.45/40,3.15/60,3.85/80,4.55/100} {
  \draw[draw=black!18] (1.0,\y) -- (8.0,\y);
  \node[font=\scriptsize, anchor=east, text=black!65] at (0.88,\y) {\lab};
}
\foreach \x/\lab in {1.75/6k,3.45/12k,5.15/24k,6.65/50k} {
  \draw[draw=black!35] (\x,1.0) -- (\x,0.92);
  \node[font=\scriptsize, text=black!70] at (\x,0.72) {\lab};
}

\draw[line, draw=black!45, dashed]
  (3.45,3.78) -- (5.15,2.24) -- (6.65,1.54);
\node[font=\scriptsize, text=black!55] at (1.75,4.35) {$\times$};
\node[tag, font=\scriptsize, anchor=west] at (1.95,4.35) {unintelligible};
\foreach \x/\y in {3.45/3.78,5.15/2.24,6.65/1.54}
  \node[marker, fill=black!45] at (\x,\y) {};

\draw[line, draw=semGreen!85!black]
  (1.75,3.22) -- (3.45,2.03) -- (5.15,1.58) -- (6.65,1.33);
\foreach \x/\y in {1.75/3.22,3.45/2.03,5.15/1.58,6.65/1.33}
  \node[marker, fill=semGreen!85!black] at (\x,\y) {};

\draw[line, draw=genOrange!90!black]
  (1.75,1.89) -- (3.45,1.46) -- (5.15,1.35) -- (6.65,1.25);
\foreach \x/\y in {1.75/1.89,3.45/1.46,5.15/1.35,6.65/1.25}
  \node[marker, fill=genOrange!90!black] at (\x,\y) {};

\draw[line, draw=black!45, dashed] (7.35,4.45) -- (7.70,4.45);
\node[marker, fill=black!45] at (7.525,4.45) {};
\node[anchor=west, font=\scriptsize] at (7.82,4.45) {AE latent only};
\draw[line, draw=semGreen!85!black] (7.35,4.10) -- (7.70,4.10);
\node[marker, fill=semGreen!85!black] at (7.525,4.10) {};
\node[anchor=west, font=\scriptsize] at (7.82,4.10) {early representation};
\draw[line, draw=genOrange!90!black] (7.35,3.75) -- (7.70,3.75);
\node[marker, fill=genOrange!90!black] at (7.525,3.75) {};
\node[anchor=west, font=\scriptsize] at (7.82,3.75) {\ours{}};

\end{tikzpicture}
}
\caption{TTS convergence with reconstruction-only and representation-based
tokenizer outputs. WER is measured from 6k to 50k downstream TTS updates.
The 6k AE-latent-only checkpoint did not produce recognizable speech, so we mark
it qualitatively rather than reporting WER. Representation-derived audio tokens
make speech emerge earlier and reduce WER faster under the same downstream
training budget.}
\label{fig:tts_convergence}
\end{figure}
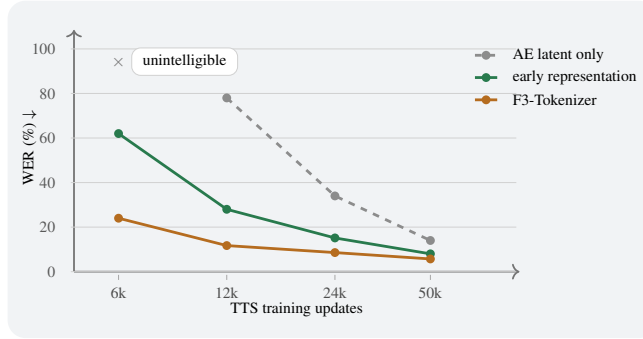

Beyond final generation metrics, we also examine whether the learned
representation makes TTS optimization easier. We compare a reconstruction-only
autoencoder output with representation-based tokenizer outputs under the same
downstream TTS setup. As shown in Figure~\ref{fig:tts_convergence},
representation-derived audio tokens make intelligible speech emerge earlier,
while the flow head still predicts the same continuous autoencoder latent
target $z$.

\FloatBarrier

\section{Discussion and Conclusion}

We presented F3-Tokenizer, a framework for adapting audio autoencoder latents
beyond reconstruction. The normalized autoencoder produces a low-dimensional,
decodable acoustic latent $z$ for reconstruction and continuous generation,
while a representation encoder maps the same latent stream into
high-dimensional representations $u$ for understanding. This separates the two
roles that are often in tension: preserving waveform-level acoustic detail and
exposing structure that is useful to an autoregressive model.

The training objectives provide complementary constraints. Reconstruction
preserves local acoustic fidelity, while channel normalization and stochastic
perturbation make the bottleneck robust without relying on a variational KL
objective. Random-quantized prediction supplies audio-only self-supervision,
frozen-LLM supervision aligns the representation with paired text, and
next-patch flow training makes the continuous acoustic target predictable from
autoregressive hidden states. Across reconstruction, probing, and generation
experiments, these constraints produce representations that remain decodable
while improving understanding and accelerating downstream TTS training.

Several limitations remain. The recipe introduces extra objectives and teacher
dependencies, so performance can depend on the coverage of random-quantized
targets, paired text, and frozen-LLM supervision. Autoencoder-only latents can
also serve as direct generation targets; future work should study whether
representation-derived audio tokens become more important for harder generation
settings such as Any2Speech~\cite{any2speech}.
Finally, the current design keeps $z$ and $u$ as separate
spaces. A unified high-dimensional representation that is both decodable and
friendly to autoregressive modeling, similar in spirit to recent
representation-autoencoder approaches in vision generation~\cite{rae,raear},
is a promising direction.

\bibliographystyle{plain}
\bibliography{references}

\end{document}